\documentstyle[aps]{revtex}
\begin{document}
\tightenlines
\title{Primordial Fluctuations of the metric
in the Warm Inflation Scenario}
\author{Mauricio Bellini\footnote{mbellini@mdp.edu.ar}}
\address{Departamento de F\'{\i}sica, Facultad de Ciencias Exactas  y
Naturales \\ Universidad Nacional de Mar del Plata, \\
Funes 3350, (7600) Mar del Plata, Buenos Aires, Argentina.}
\maketitle
\begin{abstract}
I consider a semiclassical expansion of the scalar field
in the warm inflation scenario. I study the evolution for the fluctuations
of the metric around the flat Friedmann - Robertson - Walker one.
The formalism predicts that, in a the power - law expansion of
the universe, the
fluctuations of the metric decreases with time.
\end{abstract}
\section{Introduction}
The standard inflation scenario describes a quasi-de Sitter
expansion in a supercooled scenario. 
This model separates
expansion and reheating into two distinguished time periods\cite{Linde}. 
This theory
assumes a second order phase
transition of the inflaton field, followed by a localized
mechanism that rapidly distributes the vacuum energy into
thermal energy. Reheating after inflation occurs due to particle
production by the oscillating inflaton field.

Warm Inflation is a theory of the early universe that
explains the abrupt expansion and the
exchange of energy beetwen the inflaton field
and the thermal bath. This theory was
proposed by A. Berera a few years ago\cite{1}, and
generalized in another works\cite{be}.
Quantum to classical transition\cite{2} and the power spectrum
of the primordial fluctuations was studied\cite{3} in the
framework of a stochastic approach for the warm inflation scenario. 
In this thermal scenario the rapid cooling
followed by rapid heating is replaced by a smoothened
dissipative mechanism. The warm inflation formalism
predics energy fluctuations ${\delta \rho \over \rho}$
that decreases with time in a power - law expansion,
when the universe grows very rapidly\cite{2,3}. 
In this formalism
a semiclassical expansion for the inflaton field $\varphi$
was proposed
$\varphi(\vec x,t)= \phi_c(t)+\phi(\vec x,t)$,
where $\vec x$ are the spatial cordinates, $t$ is the time,
$\phi_c(t)$ the spatially homogeneous field and $\phi(\vec x,t)$
are the quantum fluctuations.
For consistency
one requires
$<\varphi(\vec x,t)>=\phi_c(t)$, and 
$<\phi(\vec x,t)>=<\dot\phi(\vec x,t)>=0$.
In the warm inflation scenario the rapid expansion of the universe
is produced in presence of a thermal component. The kinetic
energy density ($\rho_{kin} = \rho_r + {1 \over 2} \dot\varphi^2$ )
must be small with respect to the vacuum energy density:
$\rho(\varphi)\sim 
\rho_{m}(\varphi)\sim V(\varphi) \gg \rho_{kin}$,
where $V(\varphi)$ is the potential associated with the field of
matter $\varphi$. Furthermore $\rho_m $ and $\rho_{r}= {\tau(\varphi) \over
8 H(\varphi)} \dot\varphi^2$ are the matter and radiation energy densities.
This scenario provides thermal fluctuations
compatible with de COBE data\cite{smoth} if the thermal equilibrium becomes
near the minimum of the potential $V(\varphi)$\cite{4}. Furthermore,
particles are created during the expansion of the universe, and it is
not necessary a further reheating era. The field $\varphi$ interacts
with other particles, which there are in the thermal bath at
temperature $T_r < T_{GUT}\sim 10^{15}$ GeV.

The Lagrangian density that describes the warm inflation scenario is
\begin{equation}
{\cal L}(\varphi,\varphi_{,\mu}) = - \sqrt{-g}\left[\frac{R}{16 \pi}
+\frac{1}{2} g^{\mu \nu} \varphi_{,\mu} \varphi_{,\nu}
+ V(\varphi) \right]+ {\cal L}_{int},
\end{equation}
where $R$ is the scalar curvature, $g^{\mu\nu}$ the metric
tensor, $g$ is the metric and ${\cal L}_{int}$ takes into account
the interaction of the field $\varphi$ with other fields of the
thermal bath. In this work I consider a perturbed Friedmann - Robertson -
Walker (FRW)
\begin{equation}\label{2a}
ds^2= -dt^2 + a^2(t) \  [1+h(\vec x,t)] \  d\vec x^2.
\end{equation}
Here, $a(t)$ is the scale factor and $h(\vec x,t)$ denote the
fluctuations around the FRW metric with zero global curvature
($k=0$). The perturbations $h(\vec x,t)$
are assumed to be small. Expanding $H[\varphi(\vec x, t)]$ around $\phi_c$, 
one obtains
$H(\varphi)\simeq H_c(\phi_c)+ H'(\phi_c) \  \phi(\vec x,t)$,
at first order on $\phi$. 
The perturbations of the metric, for small $\phi$, are
\begin{equation}\label{3a}
h(\vec x,t) \simeq 2 \int H_c'(\phi_c) \phi(\vec x,t) \  dt.
\end{equation}

The equation of motion for the scalar field $\varphi$ is
\begin{equation}
\ddot\varphi - \frac{1}{a^2}\nabla^2 \varphi + \left[3 H(\varphi)+
\tau(\varphi)\right] \  \dot\varphi + V'(\varphi)=0,
\end{equation}
where $\tau(\varphi) \  \dot\varphi$ describes 
the dissipation due to the interaction of the
field $\varphi$ with the fields of the thermal bath. 
We write the semiclassical Friedmann equation for a globally flat
FRW metric, which describes a globally isotropic and homogeneous
universe:
$\left<H^2(\varphi)\right> 
= \left<{8\pi\over 3}G \left(\rho_{m}+ \rho_r \right)\right>$.
Here $G=M^{-2}_p$ is the gravitational constant and $M_p$ the
Planckian mass. The matter and radiation energy densities are
$\rho_{m}(\varphi) ={\dot\varphi^2\over 2}+ \frac{1}{a^2}
\left(\vec{\nabla}\varphi\right)^2+V(\varphi)$ and
$\rho_r(\varphi) = {\tau(\varphi)\over 8 H(\varphi)} \dot\varphi^2$.

\section{The classical field $\phi_c$}

As in previous works\cite{3,4}, I consider the following relation
between the friction parameter $\tau_c$, and the Hubble one
\begin{equation}\label{qw}
\tau_c(\phi_c)=\gamma \  H_c(\phi_c),
\end{equation}
where $\gamma $ is a dimensionless constant. 
The classical field 
is defined as a solution of the equation of motion
\begin{equation}
\ddot\phi_c + [3 H_c(\phi_c)+\tau_c(\phi_c)] \  \dot\phi_c+ V'(\phi_c)=0,
\end{equation}
on the unperturbed FRW metric:
$\left< ds^2\right>  = \left<-dt^2 + a^2(t) \  [1
+h(\vec x,t)] \  d\vec x^2\right> $
$= -dt^2+ a^2(t) d\vec x^2$.

The classical Hubble parameter is
\begin{equation}
H^2_c(\phi_c)=\frac{4\pi}{3 M^2_p}\left[ \left(1+
\frac{\tau_c}{4 H_c}\right) \dot\phi^2_c+ 2 V(\phi_c)\right],
\end{equation}
and thus, the scalar potential is
\begin{equation}\label{a}
V(\phi_c) = \frac{3 M^2_p}{8\pi}\left[ H^2_c(\phi_c)-
\frac{M^2_p}{12\pi}\left(H'_c\right)^2 \left(1+\frac{\tau_c}{4 H_c}
\right)
\left(1+\frac{\tau_c}{3 H_c}\right)^{-2}\right],
\end{equation}
where one assumes that $H(\varphi)=H(\phi_c)\equiv H_c$ and
$\tau(\varphi)=\tau(\phi_c)\equiv \tau_c$. 
The eq. (\ref{a}) is obtained using the following equations
that describe the dynamics of the classical field and the 
Hubble parameter: 
$\dot\phi_c = - {M^2_p\over 4\pi}H'_c 
\left(1+ {\tau_c\over 3 H_c}
\right)^{-1}$, and
$\dot H_c  = - {M^2_p\over 4\pi} (H'_c )^2
\left(1+ {\tau_c\over 3 H_c}\right)^{-1}$,
where the prime denote the derivative with respect to the field.
With the equation for $\dot\phi_c$, one obtains the expression for the
radiation energy density
\begin{equation}
\rho_r(\phi_c)  = \frac{\tau_c}{8 H_c}\left(\frac{M^2_p}{4\pi}\right)^2
(H'_c)^2 \left(1+ \frac{\tau_c}{3 H_c}\right)^{-2}.
\end{equation}
When the thermal equilibrium holds, the temperature of the
bath is $\left<T_r  \right> \propto \left({\tau_c(\phi_c)\over
8 H_c(\phi_c)}\dot\phi^2_c\right)^{1/4}_{t\gg 1}$.
In this formalism, the expansion of the universe and the interaction 
of the inflaton with the bath are produced by the classical field
$\phi_c$. 

\section{The quantum fluctuations}

For simplicity, I consider the quantum fluctuations on
the expectation value of the metric
\begin{equation}\label{d}
\left<ds^2\right>=-dt^2+ a(t) \  d\vec x^2.
\end{equation}
However, a consistent treatment must consider the interaction of the
quantum fluctuations with the metric. Here, the simplification
$\left<H[\varphi(\vec x,t)]\right>=H_c(\phi_c)$ will be assumed.
The equation of motion for the quantum fluctuations 
[with the simplification (\ref{d})], is\cite{3,4}
\begin{equation}
\ddot\phi - \frac{1}{a^2}\nabla^2\phi + [3 H_c + \tau_c ] \dot\phi 
+ V''(\phi_c) \phi =0.
\end{equation}
The structure of this equation 
can be simplified with the mapp $\chi = e^{3/2 \int(H_c
+\tau_c/3) dt} \  \phi $, and one obtains
\begin{equation}
\ddot\chi - a^{-2} \nabla^2 \chi - \frac{k^2_o}{a^2}\chi =0,
\end{equation}
where 
$k^2_o(t)$ $= a^2 \left[ {9\over 4}\left(H_c+ {\tau_c\over 3}\right)^2-
V''(\phi_c)+{3\over 2}\left( \dot H_c + {\dot\tau_c\over 3}\right)
\right]$.
The field that describes the quantum fluctuations can be written as
a Fourier expansion of the modes $\xi_k(t) e^{i \vec k. \vec x}$
\begin{equation}\label{f}
\chi(\vec x,t)=\frac{1}{(2\pi)^{3/2}}\int
d^3k \  \left[ a_k e^{i \vec k. \vec x} \xi_k(t)+h.c.\right].
\end{equation}
Since in the metric (\ref{d}) there are not taken into account
the quantum fluctuations, the field (\ref{f}) can be consider as free.
However, in a more consistent formalism must be considered
the interaction of the quantum fluctuations with
the fluctuations of the metric. 
I denote with
$a^{\dagger}_k$ and $a_k$, 
the creation and annihilation operators.
These operators satisfy the commutation relations
$\left[a_k,a^{\dagger}_{k'}\right]$ $=  \delta^{(3)} (\vec k - \vec k')$ and
$\left[a_k,a_{k'}\right] $ $=  \left[a^{\dagger}_k,a^{\dagger}_{k'}
\right] = 0$.
The commutation relation between the operators $\dot\chi$ and $\chi$ is
\begin{equation}\label{com}
\left[ \chi(\vec x,t), \dot\chi(\vec x,t)\right]=i \delta (\vec x-\vec x'),
\end{equation}
which is satisfied for
$\xi_k \dot\xi^*_k - \dot\xi_k \xi^*_k = i$.

I am interested in the study of the universe on a scale greater than
the observable universe. Thus, I consider the 
field $\chi$, but taking into account only 
the modes with wavelength of size
$l \ge \frac{1}{\epsilon k_o}$,
where $\epsilon \ll 1$ is a dimensionless constant.
This field is given by
\begin{equation}\label{t}
\chi_{cg}(\vec x,t)= \frac{1}{(2\pi)^{3/2}} \int
d^3k \  \theta(\epsilon k_o - k) \  \left[ a_k e^{i \vec k.\vec x}
\xi_k + h.c.\right].
\end{equation}
The coarse - grained field (\ref{t}) satisfy the following
stochastic equation
\begin{equation}\label{h}
\ddot\chi_{cg} - \frac{k^2_o}{a^2}\chi_{cg}=
\epsilon \left( \frac{d}{dt}\left( \dot k_o \eta\right)+
2 \dot k_o \kappa\right),
\end{equation}
with the noises
\begin{eqnarray}
\eta(\vec x,t) &=& \frac{1}{(2\pi)^{3/2}}
\int d^3k \  \delta(\epsilon k_o-k)
\left[ a_k e^{i \vec k.\vec x}\xi_k + h.c.\right], \nonumber \\
\kappa(\vec x,t) &=& \frac{1}{(2\pi)^{3/2}}
\int d^3k \  \delta(\epsilon k_o-k)
\left[ a_k e^{i \vec k.\vec x}\dot\xi_k + h.c.\right]. \nonumber
\end{eqnarray}
The equation (\ref{h}) is an second order
operatorial stochastic equation, and
also can be obtained in the standard inflation formalism
(for $V(\varphi)=0$)\cite{BCMS}.

Complex to real transition of the modes $\xi_k$ was studied
in a previous article\cite{4}. When this transition occurs
one obtains
$\xi_k(t)\xi^*_k(t')\simeq \xi_k(t) \xi(t')$,
and the commutators
$[\chi_{cg}, \eta]$, $[\chi_{cg}, \kappa]$, $[\eta, \kappa]$, 
and (\ref{com}) are null.
Hence, the self - correlation of the coarse - grained field
$\chi_{cg}$ becomes
\begin{equation}\label{u}
\left< \chi_{cg}(t)\chi_{cg}(t') \right>=\frac{1}{(2\pi^{3/2})}
\int^{\epsilon k_o(t')}_{\epsilon k_o(t)}
d^3k \  \xi_k(t) \  \xi_k(t').
\end{equation}
The radiation energy fluctuations are
$\delta \rho_r(t) =  \left| 2 \left({\gamma\over 8}\right)
\left(1+{\gamma\over 3}
\right)^{-2} H'_c H''_c \right| \left<\phi^2_{cg}\right>^{1/2}$,
and the radiation energy density is
$\rho_r(t)$ $= \left({\gamma\over 8}\right)\left(1+{\gamma\over 3}
\right)^{-2}\left(H'_c\right)^2$.
When the radiation energy fluctuations
${\delta \rho_r \over \rho_r}$,
decreases with time,
the thermal equilibrium holds for sufficiently large times.
In this case, the 
Fourier transform of (\ref{u}) gives the spectral density
for the quantum fluctuations in the infrared sector
\begin{equation}
S[\chi_{cg}; \omega_k] =  \frac{1}{\pi} \int^{\infty}_{0}
dt'' \  cos[\omega t''] 
\int^{\epsilon k_o(t+t'')}_{\epsilon k_o(t)}
dk \  k^2 \  \xi_k(t)\xi_k(t+t'').
\end{equation}
Here, $\omega_k$ is the frecuency of oscillation for each mode with
wavenumber $k$.
In a semiclassical representation to the 
warm inflation scenario, the fluctuations of the metric are dues
to the quantum fluctuations $\phi_{cg} =
e^{-3/2 \int (H_c+\tau_c/3) dt} \  \chi_{cg}$, on a
scale much greater than the scale of the observable universe.
The perturbed metric for the FRW universe is given by eq. (\ref{2a}),
with $h(\vec x, t)$ given by eq. (\ref{3a}). The temporal evolution
of $h(\vec x,t)$ is given by
\begin{equation}\label{r}
\left< h^2(t)\right>^{1/2}\simeq a^{-2} \left[ 2 \int
H'(\phi_c) \  <\phi_{cg}^2(\vec x, t)>^{1/2} \  dt\right].
\end{equation}
Quantum to classical transition of the perturbations of the
metric holds due to the quantum to classical transition of the
coarse - grained quantum field $\phi_{cg}$. 
It is produced when all of the modes of $\chi_{cg}$
become real, i.e.\cite{4}:
${1\over N(t)} \sum^{k=\epsilon k_o}_{k=0} \alpha_k(t)\ll 1$.
Here $N(t)$ is the time dependent number of degrees of freedom
of the infrared sector, which increases with time.
The function $\alpha_k(t)$ was defined in a previous work\cite{3}, and is:
$\alpha_k = \left| {v_k(t)\over u_k(t)}\right|$,
for $\xi_k(t)= u_k(t)+ i \  v_k(t)$. This effect is due to the
time evolution of the superhorizon with size $l(t) \ge {1\over \epsilon
k_o}$. Thus, quantum to classical transition of the coarse - grained
field occurs when $\alpha_{k=\epsilon k_o} \rightarrow 0$.

\section{An example: fluctuations of the metric in
power - law inflation}.

In this section I estime the fluctuations of the metric
in the the example for power - law inflation.
Here, the scale factor and the Hubble parameter are
$a(t)  = H^{-1}_o (t/t_o)^p$, and $H_c(t)= \frac{p}{t}$, respectively.
The temporal evolution of the scalar field is
$\phi_c(t)=\phi_o - m \  {\rm ln}\left[\frac{H_o}{p}t\right]$.
Hence, the scalar potential $V(\phi_c)$ is given by
\begin{equation}
V(\phi_c) =   \frac{3 M^2_p H^2_o}{8 \pi} e^{2 \phi_c/m} 
\left[ 1 - \frac{M^2_p }{12 \pi m^2} \left(1+\frac{\gamma}{4}\right)
\left(1+\frac{\gamma}{3}\right)^{-2}\right].
\end{equation}
The temporal evolution of the quantum fluctuations is
\begin{equation}
\left.<\phi^2_{cg}(t)>^{1/2} \right|_{t\gg 1} \propto
t^{2\nu
p(1-\nu/p-p/\nu)+3/2-\gamma/2},
\end{equation}
with
$\nu = {\sqrt{1+9 p^2
(1+\gamma)^2- 6p [(1+\gamma)+1]
+9 (1+\frac{\gamma}{4})(1+\frac{\gamma}{3})^{-2}}\over2 (p-1)}$.
These fluctuations decrease with time when the 
universe grows very rapidly (for $p\gg 1$). So, the 
fluctuations of the metric
are given (as $H'_c(t) \sim t^{-1}$) by
\begin{equation}
\left< h^2(t)\right>^{1/2} \sim t^{2\nu p(1-\nu/p-p/\nu)+3/2-\gamma/2},
\end{equation}
which, since $<\phi^{2}_{cg}>^{1/2}$, decreases with time for 
$p \gg 1$.
The power spectral density for the matter field fluctuations is 
$S\left[ \chi_{cg},\omega_k={2 k \over t^p}\right] \propto
\left|\omega_k\right|^n$, with $n = -[4\nu(p-1)-2p+4]$.
The fluctuations of radiation energy density are
\begin{equation}
\left.\frac{\delta \rho_r}{\rho_r}\right|_{t\gg 1} \sim
t^{2 \nu p(1-\nu/p-(5/4)p/\nu)+1/2-\gamma/2},
\end{equation}
which decreases for $p$ sufficiently large. Thus,
for $p \gg 1$ the thermal equilibrium
holds due to ${d  [ \delta \rho_r / \rho_r] \over d t} < 0$.

\section{Final Comments}

In this work was studied the evolution for the fluctuations of the metric
with a semiclassical representation of the inflaton field $\varphi$,
in the warm inflation scenario. The mean temperature is smaller than
the GUT temperature (i.e., $\left< T_r \right> \  < \  T_{GUT} \simeq
10^{15}$ GeV). In this theory the classical matter field lead to
the expansion of the universe, while the fluctuations of the matter field
generate the inhomogeneities of the metric, and thus local curvature
of the spacetime. However the expectation value of the
curvature is zero, in consistency with a globally
flat perturbed Friedmann -
Robertson - Walker metric here considered.
In this framework the quantum fluctuations averaged over
a scale much bigger than the observable universe 
are responsible (not only of the
radiation energy density fluctuations, thermal fluctuations and
matter energy density fluctuations - these topics
were studied in another 
previous works\cite{2,3,4}) for the fluctuations of the metric.
This calculation was done with the assumption that the
quantum fluctuations are very small. Thus, was possible
to expand the coarse - grained field as free. In a more
appropiate treatment for the coarse - grained
field, it must be consider as interactuant
with the metric. 

In the example here considered I observe that the fluctuations
of the metric decreases with time,
in a power - law expansion when the expansion is very rapid.
Furthermore, the radiation energy fluctuations
decreases with time for a
sufficiently large rate of expansion of the universe.
Hence, when the universe expands very rapidly, the thermal equilibrium
holds.

\end{document}